\documentclass[aps,prb,twocolumn,superscriptaddress]{revtex4}
\usepackage[normalem]{ulem}
\usepackage{float}
\usepackage[usenames]{color}
\usepackage{graphicx}
\usepackage{ulem}
\usepackage{amsfonts}
\usepackage{amsmath}
\usepackage{natbib}

\newcommand{\del}[1]{{  }}
\renewcommand{\P}{p}

\newcommand{\G}{\mathcal{G}}

\usepackage{color}
\newcommand\ta[1]{\bgroup\color{blue}\bfseries{#1}\egroup}
\newcommand\tr[1]{\bgroup\color{red}\bfseries{}\egroup}

\def\TM{\mathbf{T}}

\begin{document}

\title{Criticality and isostaticity in fiber networks}

\author{Chase P. Broedersz}
\affiliation{Department of Physics and Astronomy, Vrije Universiteit, Amsterdam, The Netherlands}
\author{Xiaoming Mao}
\affiliation{Department of Physics and Astronomy, University of Pennsylvania, Philadelphia, Pennsylvania 19104, USA}
\author{T.C. Lubensky}
\affiliation{Department of Physics and Astronomy, University of Pennsylvania, Philadelphia, Pennsylvania 19104, USA}
\author{F.C. MacKintosh}
\affiliation{Department of Physics and Astronomy, Vrije Universiteit, Amsterdam, The Netherlands}

\date{\today}

\maketitle \noindent\textbf{The rigidity of elastic networks depends sensitively on their internal connectivity 
and the nature of the interactions between constituents. 
Particles interacting via central forces undergo a zero-temperature rigidity-percolation transition near the isostatic threshold, 
where the constraints and internal degrees of freedom are equal in number~\cite{Maxwell1864,Thorpe1983}. 
Fibrous networks, such as those that form the cellular cytoskeleton~\cite{Bausch2006,Fletcher2010}, 
become rigid at a lower threshold due to additional bending constraints.
However, the degree to which bending governs network mechanics remains a subject of considerable debate~\cite{Head2003,Wilhelm2003,Gardel2004,Storm2005,Onck2005,Heussinger2006,Chaudhuri2007,Buxton2007}. 
We study disordered fibrous networks with variable coordination number, 
both above and below the central-force isostatic point.
This point controls a broad 
crossover from stretching- to bending-dominated elasticity.
Strikingly, this crossover exhibits an anomalous power-law dependence of 
the shear modulus on \emph{both} stretching and bending rigidities.
At the central-force isostatic point---well above the rigidity 
threshold---we find divergent strain fluctuations together with a 
divergent correlation length $\xi$, implying a breakdown of continuum 
elasticity in this simple mechanical system on length scales less than 
$\xi$.}

\begin{figure}
\begin{center}%\vspace{-0.1in}
\includegraphics[width=\columnwidth]{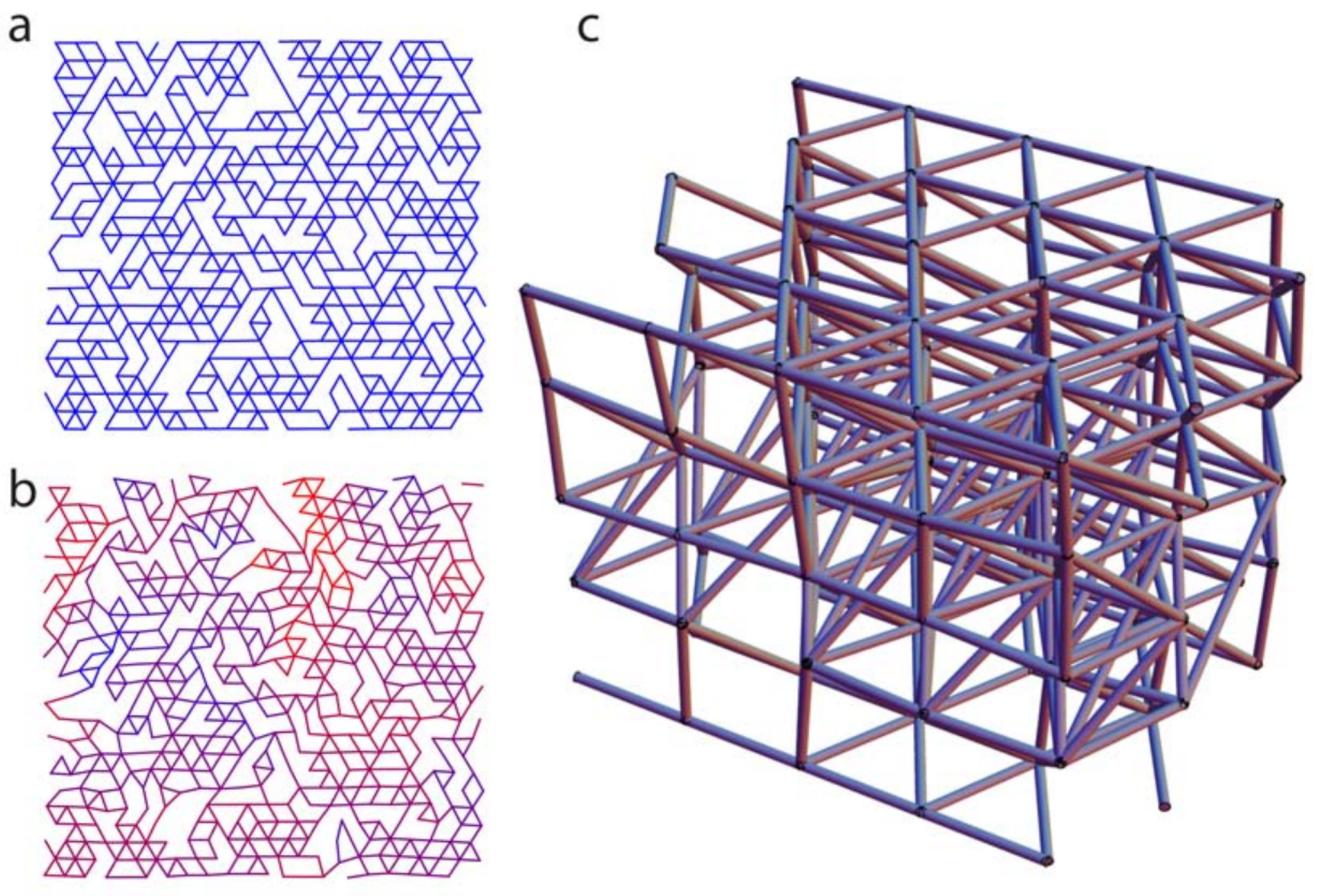}
\caption{\label{fig:networkdiagrams} \textbf{Fiber networks arranged on 2D and 3D lattices} A small section of a sheared diluted
triangular network near isostaticity with relatively stiff filaments ($\kappa=10^{1}$
in units of $\mu\ell_0^2$) (a) and floppy filaments ($\kappa=10^{-5}$) (b).
The deviation of the local deformation from a uniform deformation is indicated
by color, where blue corresponds to a uniform or affine deformation and red
corresponds to a highly non-affine deformation.  (c) An example of a small
section of the diluted FCC network at $\P=0.7$. To probe the mechanical properties
of this network we shear the 111-plane (shown on top) along the direction of
one of the bond angles in this plane.}
\end{center}\vspace{-0.2in}
\end{figure}

There are numerous examples of stiff-fiber networks, ranging from carbon nanotube gels at the small scale to felt and paper at the macroscopic scale~\cite{Hall2008,Hough2004,Kabla2007}.
In addition, critical biological components such as the intra-cellular cytoskeleton and extra-cellular matrices of collagen and fibrin are composed of such networks~\cite{Pedersen2005}. Here, we use a combination of simulations and effective medium theory we study the elasticity of disordered fiber networks composed of straight, stiff filaments organized
on a triangular lattice in 2D and face centered cubic (FCC) lattice in
3D, as illustrated in Fig.~\ref{fig:networkdiagrams}. Undiluted, these networks have a coordination number $z_m=6$
(triangular lattice) and $z_m=12$ (FCC), placing them well above the
central-force (CF) isostatic point $z_c=2d$ in $d$
dimensions~\cite{Maxwell1864, Wyart2008a}.  We explore the
effects of network connectivity---both above and below $z_c$---by
removing filament segments between vertices with a probability $1-\P$.

\begin{figure*}
\begin{center}%\vspace{-0.1in}
\includegraphics[width=2\columnwidth]{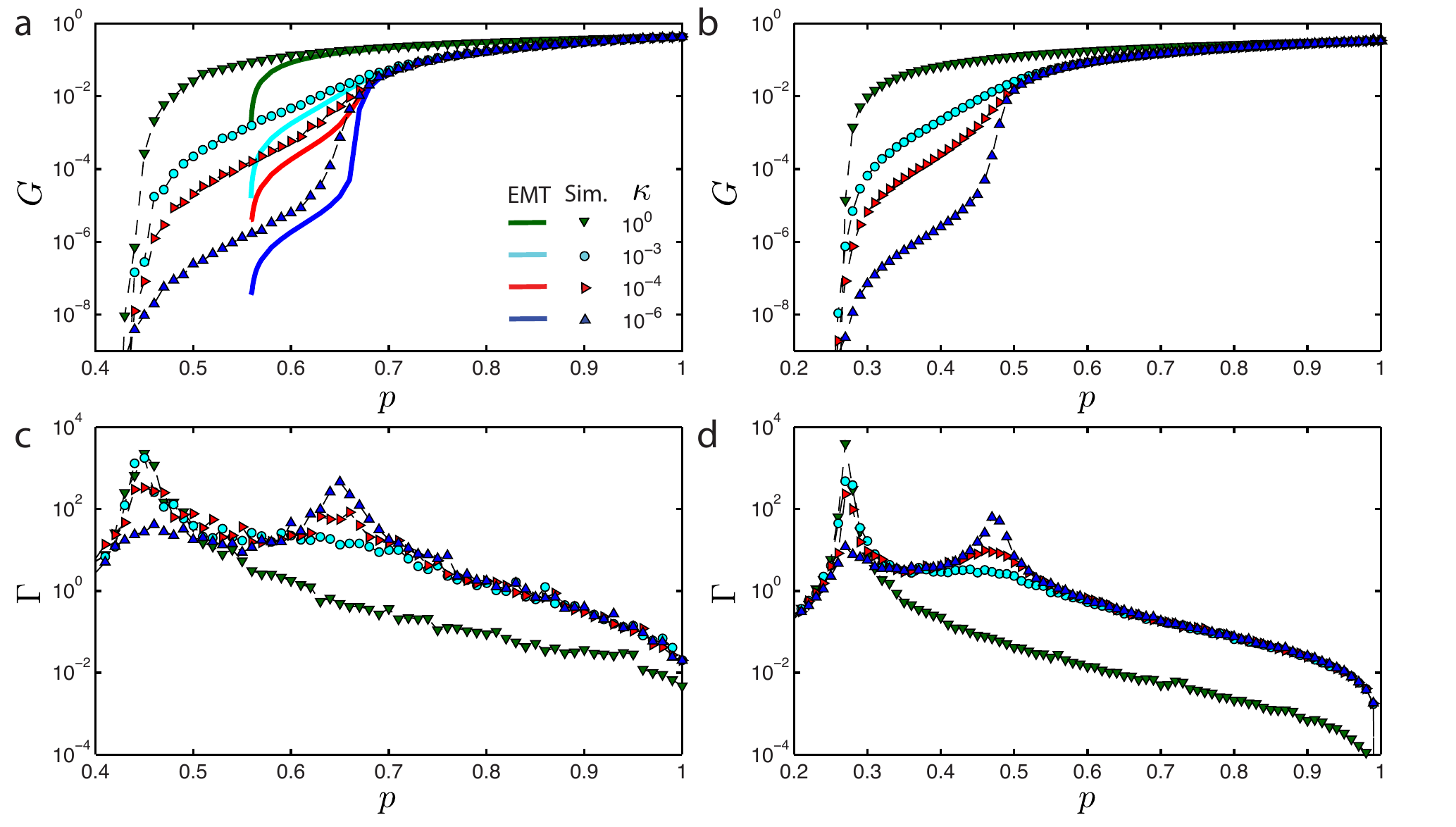}
\caption{\label{fig:gandgamma} \textbf{Mechanics and non-affine strain fluctuations} The shear modulus $G$ in units
of $\mu/\ell_0$ as a function of $\P$ for a range of filament
bending rigidities $\kappa$ for the 2D triangular lattice (a)
and the 3D FCC lattice (b).  The EMT calculations for the 2D
triangular lattice are shown as solid lines. The non-affinity
measure $\Gamma$ is shown as a function of $\P$ for various
values of $\kappa$ for the 2D triangular lattice (c) and the 3D
FCC lattice (d). The values of $\kappa$ in units of
$\mu\ell_0^2$ are  $10^{0}$ (green), $10^{-3}$ (cyan),
$10^{-4}$ (red) and $10^{-6}$ (blue).}
\end{center}\vspace{-0.2in}
\end{figure*}

The mechanical response of the fibers in the network is
determined by their bending rigidity $\kappa$ and stretching
modulus $\mu$. For small deformations, the stretching and
bending energy of the network can be expanded to quadratic
order in the displacements ${\bf u}_i$ from the undeformed
reference state at each vertex $i$,
\begin{eqnarray}
\label{eq:stretchenergy} E_{\rm stretch}&=&\frac{1}{2}\frac{\mu}{\ell_0} \sum_{\langle ij \rangle} g_{ij} \left({\bf u}_{ij} \cdot {\bf \hat{r}}_{ij}  \right)^2\\
\label{eq:bendenergy}
E_{\rm bend}&=&\frac{1}{2}\frac{\kappa}{\ell_0^{3}} \sum_{\langle ijk \rangle} g_{ij}g_{jk} \left[\left( {\bf u}_{jk}-{\bf u}_{ij}\right)\times {\bf \hat{r}}_{ij}\right]^2,
\end{eqnarray}
where $\ell_0$ is the lattice spacing, ${\bf u}_{ij}={\bf u}_{j}-{\bf
u}_{i}$ and ${\bf \hat{r}}_{ij}$ is the unit vector oriented  along the
$ij$-th bond in the undeformed reference state. Here, $g_{ij}=1$ for
present bonds and $g_{ij}=0$ for removed bonds. The summation extends
over neighboring pairs of vertices in the stretching term
[Eq.~(\ref{eq:stretchenergy})], and  over coaxial neighboring bonds in
the bending term [Eq.~(\ref{eq:bendenergy})]. Thus, in our networks the 
cross-links at each vertex are freely hinging. 

To investigate the mechanical response of a network, we calculate its
shear modulus $G$ numerically. The diluted networks exhibit a finite
shear modulus well below the CF isostatic point (expected at $\P_c=2/3$
in 2D and  $\P_c=1/2$ in 3D), as shown in Fig.~\ref{fig:gandgamma}a,b;
$G$ vanishes at a $\kappa$-independent rigidity percolation point
located at $\P_b = 0.445 \pm 0.005$ (2D triangular lattice) and $\P_b =
0.275\pm 0.005$ (3D FCC lattice),  consistent with a floppy mode
counting argument that includes the bending constraints (suppl. info.).
Plots of $G$ versus $\P$ for different $\kappa$ are shown for the
triangular and FCC lattices in Figs.~\ref{fig:gandgamma} a,b. 
For $\P>\P_c$, $G$ approaches a nearly $\kappa$-independent
stretching dominated limit with $G \sim \mu$. In contrast, for $\P<\P_c$, $G$ falls
off reaching a bending dominated limit with $G \sim \kappa$ as $\P \to
\P_b$. The most interesting behavior occurs near $\P_c$ as a function
of $\kappa$.  There is a stretch dominated regime at large $\kappa$ and
bending dominated one at small $\kappa$ with a broad intermediate crossover
regime with $G$ depending on both $\kappa$ and $\mu$.

\begin{figure*}
\begin{center}%\vspace{-0.1in}
\includegraphics[width=2 \columnwidth]{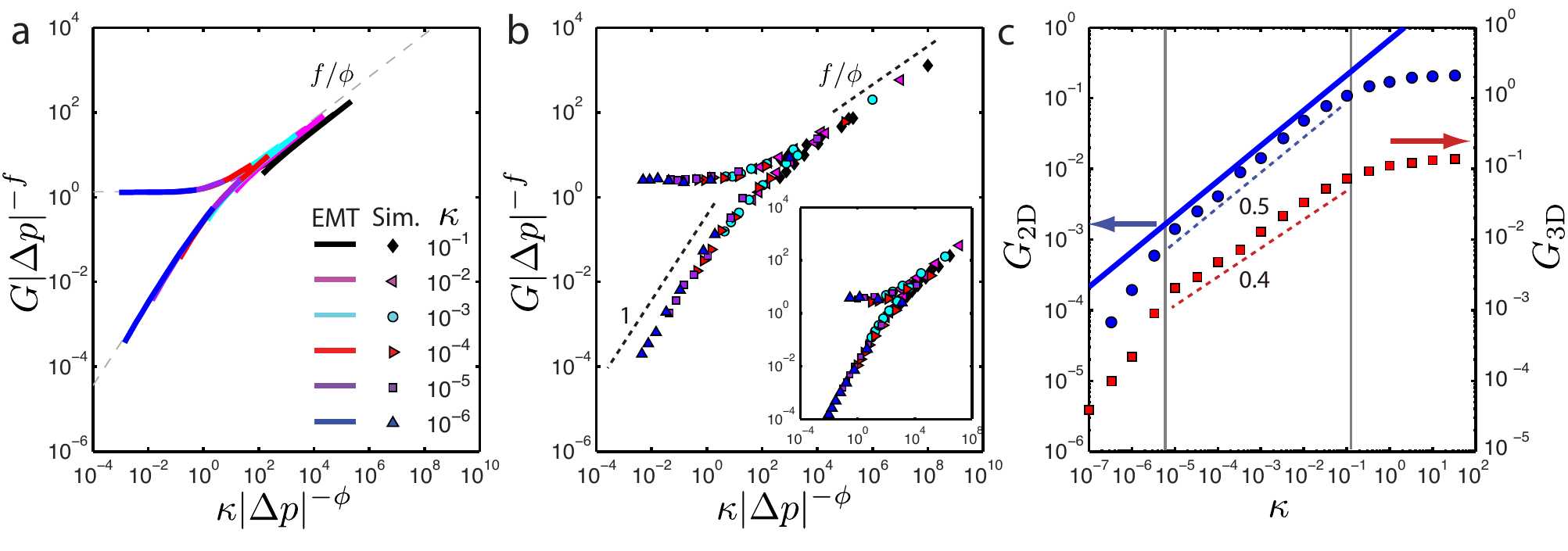}
\caption{\label{fig:widom} \textbf{Scaling analysis of the mechanics and anomalous elasticity} Scaling of the shear modulus in the vicinity
of the isostatic point with the scaling form $G |\Delta \P|^{-f}=
\G_{\pm} \left(\kappa |\Delta \P|^{-\phi} \right)$, with $G$ in units
of $\mu/\ell_0$, for the mechanical properties of the diluted
triangular lattice for the EMT calculations (a) and the simulations (b)
for a broad range of filament bending rigidities  ($\kappa$ in units of
$\mu\ell_0^2$: $10^{-1}$ black, $10^{-2}$ magenta,
$10^{-3}$ cyan, $10^{-4}$ red, $10^{-5}$ purple and $10^{-6}$ blue).
The asymptotic form of the scaling function for low $\kappa$ is shown
as a dashed grey line in (a). The scaling for the numerical data on the
3D FCC lattice is shown as an inset in b. The EMT exponents are $f_{\rm
EMT}=1$, $\phi_{\rm EMT}=2$. In contrast, for the numerical data we
obtain $f=1.4\pm0.1$, $\phi=3.0\pm0.2$ (2D) and $f=1.6\pm0.2$
$\phi=3.6\pm0.3$ (3D).  The scaling for the numerical data is performed
with respect to the isostatic point of the finite system for which we
find $\P_c(W) \approx 0.651$ (2D, $W$=200) and $\P_c(W) \approx 0.473$
(3D, $W$=30). (c) The shear modulus as a function of $\kappa$ close to
the isostatic point for the triangular lattice ($\P=0.643$, blue
circles) and the FCC lattice ($\P = 0.47$, red squares). At low
$\kappa$ there is a bending dominated regime $G_{\rm bend}\sim\kappa$,
at intermediate $\kappa$ there is a regime in which stretching and
bending modes couple strongly with $G\sim\mu^{1-x}\kappa^x$, where
$x=0.50\pm0.01$ (2D) and $x\approx0.40\pm0.01$ (3D). The EMT
calculation for $\kappa/\mu\gg\vert\Delta\P\vert^{\phi_{\textrm{EMT}}}$  is shown as a solid blue line.}
\end{center}\vspace{-0.2in}
\end{figure*}

To gain insight into the mechanical behavior of our models, we
developed a new effective medium theory (EMT) or coherent potential
approximation (CPA)~\cite{Soven1969,Feng1985,Das2007} for lattices with
bending forces, which we discuss in more detail in the methods section,
whose results for $G$ for different $\kappa$ are shown in
Fig.~\ref{fig:gandgamma}a. These results overestimate the rigidity
percolation point $\P_b$. Nonetheless, this model captures the
essential features of the simulations well, including the crossover
between stretching and bending dominated regimes close to $\P_c$. Our
EMT theory predicts that when $\kappa/\mu \ll \Delta \P$, $G$ can be
expressed in the vicinity of $\P_c$ in the scaling form
\begin{equation}
\label{eq:widomscale} G=\mu |\Delta \P|^f \G_{\pm} \Big(\frac{\kappa}{\mu} |\Delta \P|^{-\phi} \Big),
\end{equation}
where $f=f_{\rm EMT} = 1$ and $\phi= \phi_{\rm EMT} = 2$ are, respectively, the
rigidity and crossover critical exponents.  This scaling form is
analogous to that for the conductivity of a random resistor
network~\cite{Straley1976} with bonds occupied with resistors of
conductance $\sigma_{>}$ and $\sigma_<$ with respective probabilities
$\P$ and $(1-\P)$.  When $y\ll 1$, $\G_+(y) \sim \text{const.}$ and $\G_-(y)
\sim y$ implying $G \sim \mu|\Delta \P|^f$ for $\Delta p >0$ and $G
\sim \kappa |\Delta \P|^{f-\phi}$ for $\Delta p <0$. In the opposite
limit $(\kappa/\mu)\gg |\Delta \P|^{\phi}$, $G$ must become independent of
$\Delta \P$ since it is neither zero nor infinite at $\Delta \P = 0$.
Equation (\ref{eq:widomscale}) predicts $G \sim \kappa^{f/\phi}
\mu^{1-(f/\phi)}$, which reduces of $G\sim \kappa^{1/2} \mu^{1/2}$ in
the EMT theory.  The full EMT results for $G$ along with the scaling
form valid at $\kappa/\mu \ll \vert\Delta p\vert^{\phi}$ are shown in
Fig.~\ref{fig:widom}a.

Our simulation data for both 2D and 3D networks are well described by the scaling hypothesis in Eq.\ (\ref{eq:widomscale}), consistent with a second-order transition for $\kappa=0$ in both cases~\cite{Chubynsky2007}.
Fig.~\ref{eq:widomscale}b shows the
results for both the triangular case and FCC cases (inset).
As expected from previous simulation work~\cite{Head2003,Wilhelm2003,Buxton2007}, we find
a bending-dominated regime at small $\kappa$ and a stretching-dominated
regime at large $\kappa$. Consistent with the EMT prediction above, we find a previously unexpected 
intermediate regime with $G\sim
\kappa^x\mu^{1-x}$ where $x=f/\phi
\approx0.50\pm 0.01$ (2D) and $f/\phi \approx0.40\pm 0.01$ (3D). These results are
consistent with our exponents obtained above (Table 1, suppl. info.). While the
extent of this intermediate regime is bounded from above by the
affine modulus, it can extend to arbitrarily small $\kappa>0$ as the
system is brought closer to CF isostaticity.

To investigate the nature of the various mechanical regimes, we examine
the local deformation field in our simulations. Several methods have
been proposed to quantify the deviation from a uniform (affine) strain
field~\cite{Head2003, DiDonna2005, Liu2007}. Here we utilize a measure for this
non-affinity given by
\begin{equation}\label{eq:NA}
\Gamma=\frac{1}{N \gamma^2}\sum_i \big\lbrack{\bf u}_i-\textbf{u}_i^{({\rm aff})}\big\rbrack^2,
\end{equation}
where $\textbf{u}_i^{({\rm aff})}$ is the affine displacement of vertex
$i$ and $N$ is the number of vertices. This quantity varies over eight
orders of magnitude, indicating non-affine fluctuations that depend
strongly on both $\kappa$ and $\P$, as shown in
Figs.~\ref{fig:gandgamma}c,d. For stretch-dominated networks (high
$\kappa$), we find a monotonic increase in non-affine fluctuations
with decreasing $\P$, which appear to diverge at $\P_b$. Remarkably,
for smaller values of $\kappa$, a second peak in $\Gamma$ develops at
$\P_c$. Importantly, the development of this peak coincides with the
appearance of a crossover between the stretching and bending regimes
(Figs.~\ref{fig:gandgamma}a-d).

\begin{figure*}
\begin{center}%\vspace{-0.1in}
\includegraphics[width=2 \columnwidth]{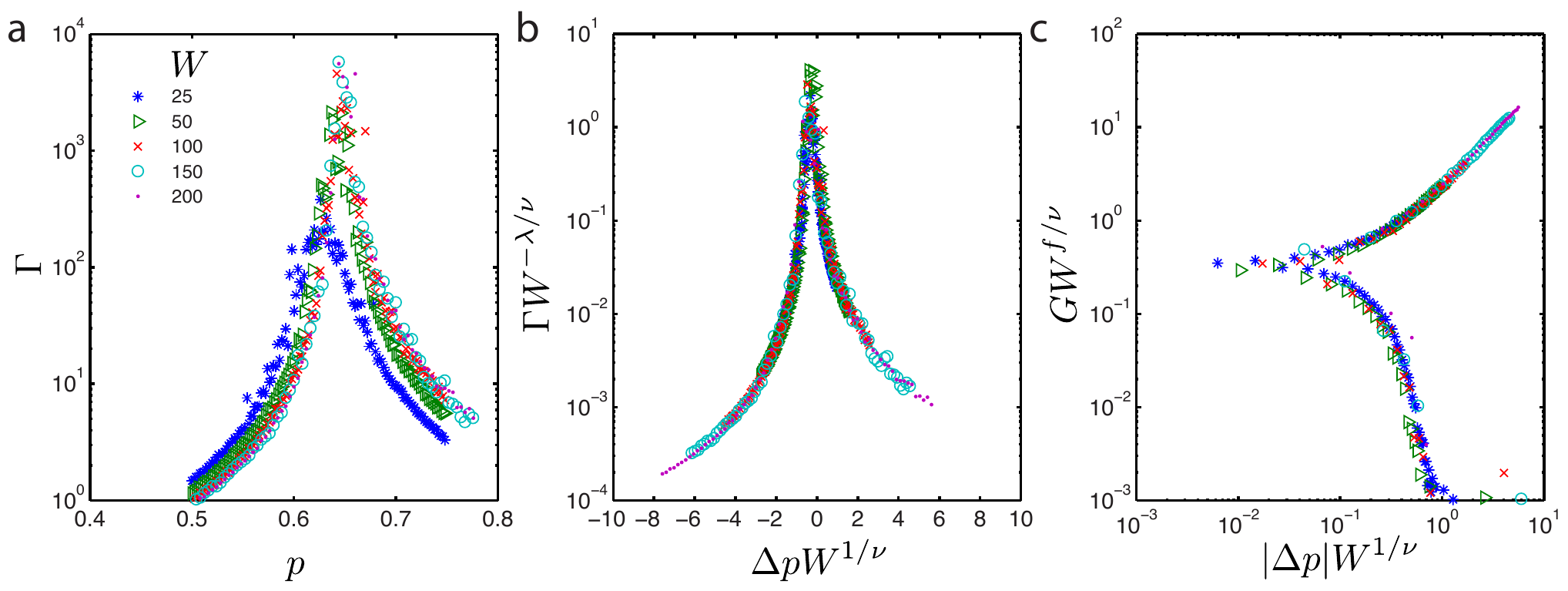}
\caption{\label{fig:finitesize} \textbf{Finite size scaling} (a) The non-affinity measure
$\Gamma$ for the 2D triangular lattice at $\kappa=0$ for
various systems sizes $W$ (25 blue, 50 green, 100 red, 150 cyan
and 200 purple).  Finite size scaling of the non-affinity
measure $\Gamma$ according to the scaling form $\Gamma
=W^{\lambda/\nu} \mathcal{F}_{\Gamma,\pm} \big(\Delta \P
W^{1/\nu}\big)$ (b) and of the shear modulus with the scaling
form $G =W^{-f/\nu} \mathcal{F}_{G,\pm} \big(|\Delta \P|
W^{1/\nu}\big)$ (c). Here $\Delta \P=\P-\P_c$, where
$\P_c=0.659\pm0.002$. The exponents we obtain are
$\lambda/\nu=1.6\pm0.2$, $\nu=1.4\pm0.2$ and $f/\nu=0.9\pm
0.1$.}
\end{center}\vspace{-0.2in}
\end{figure*}

The critical behavior we observe suggests both a divergence of the
non-affine fluctuations according to~\cite{Wyart2008a} $\Gamma= \Gamma_{\pm} |\Delta
\P|^{-\lambda}$ and the existence of a divergent
length-scale $\xi=\xi_{\pm} |\Delta \P|^{-\nu}$ near the critical point
$P_c$ for vanishing $\kappa$. However, the divergence of $\xi$ is
limited by the system size $W$, which should suppress the divergence
of $\Gamma$. Consistent with this picture, we find that the the
location of the cusp in the local fluctuations $\Gamma$ shift towards
higher $\P$ with increasing $W$ according to $\P_c(W)=\P_c +b
W^{-1/\nu}$, with $\nu=1.4\pm0.2$ and $\P_c=0.659\pm0.002$ (suppl.
info.); these values are consist with previous reports on generic
CF networks~\cite{Jacobs1996}. In addition, the amplitude of
$\Gamma$ increases with system size (Fig.~\ref{fig:finitesize}a), in
quantitative accord with the expected finite-size scaling.
Specifically, we find a good collapse of the simulation data with
$\Gamma =W^{\lambda/\nu} \mathcal{F}_{\Gamma,\pm} \left(\vert\Delta
\P\vert W^{1/\nu}\right)$ over a range of system sizes, with
$\lambda/\nu=1.6\pm0.2$ and $\nu=1.4\pm0.2$, as shown in
Fig.~\ref{fig:finitesize}b. Similarly, the shear modulus exhibits
finite-size scaling (Suppl. info.)  according to  $G =W^{-f/\nu}
\mathcal{F}_{G,\pm} \left(|\Delta \P| W^{1/\nu}\right)$, as shown in
Fig.~\ref{fig:finitesize}c. We obtain a good collapse of the elasticity
data using $f/\nu=0.9 \pm 0.1$, along with $\nu$ determined from the
finite-size scaling of $\Gamma$ (Fig.~\ref{fig:finitesize}b and suppl.
info), consistent with the value of $f$ obtained from the scaling in
Fig.~\ref{fig:widom}. Thus, we find a scale-dependence of the shear
modulus that is consistent with critical behavior governed by the CF
isostatic critical point. Furthermore, the critical behavior in these
purely mechanical networks is accompanied by shear-induced divergent
non-affine fluctuations. These results imply a breakdown of continuum elasticity
below the divergent length-scale $\xi$.

The undiluted triangular and FCC lattices we study have an average
coordination number greater than $2d$ and thus are above the Maxwell
central-force isostatic threshold. These networks also consist of
infinitely long filaments. Cutting bonds as we do introduces
both finite length polymers, as well as lower connectivity, down to the 
CF threshold and below. Cytoskeletal and extracellular networks 
can have $z$ as low as 3 (e.g., in branched networks) and as high as 6 (in the 
case of actin-spectrin networks), although they typically have a 
local connectivity $z\simeq4$, where two filaments are
connected by a cross-link. 
As a consequence, the CF isostatic point is expected to occur for high molecular weight in 2D.
We conjecture that there is an analogous
crossover behavior for such networks, including the anomalous scaling behavior for the elasticity. 
In addition, we expect that our results for the crossover behavior will apply to
bond-bending models on similar lattices to ours~\cite{Thorpe1983,Schwartz1985,He1985,Sahimi1993}
for rigidity percolation and network glasses that include bending
forces between bonds pairs at each network node.
Finally, from the perspective of critical phenomena more generally, 
the kind of crossover behavior we find here is in contrast to most thermal systems, 
where a field or coupling constant leads to a crossover from one critical system to another, 
such as from the Heisenberg model to the Ising model~\cite{Fisher83}. In such systems, 
there is a continuous evolution of the critical point that is governed by the
crossover exponent $\phi$. Interestingly, we find no such continuous evolution, but rather a 
discontinuous jump in the critical point $p_c$ as soon as $\kappa$ becomes nonzero.

\section{Methods}

\noindent \emph{Simulations} The mechanical response of the
network is determined in our simulations by applying a shear
deformation with a strain $\gamma$. This is realized by
translating the horizontal boundaries to which the filaments
are attached, after which the internal degrees of freedom  are
relaxed by minimizing the energy using a conjugate gradient
algorithm~\cite{Vetterling2002}. To reduce edge effects in our
simulation, periodic boundary conditions are employed at all
boundaries. The shear modulus of the network is related to the
elastic energy through $G=\frac{2}{V_0 W^d}\frac{E}{\gamma^2}$
for a small strain $\gamma$, where $V_0$ is the area/volume of
a unitcel. Here $W^d$ is the system size, which in our
simulations is  $W^2\approx 40000$ (2D) and $W^3\approx  30000$
(3D), and we used strains no larger than $\gamma=0.05$.

\noindent \emph{EMT} The EMT maps the diluted random network to an
undiluted uniform effective medium (EM) with respective stretching
modulus $\mu_m$ and bending rigidity $\kappa_m$, which are determined
self-consistently as follows.  In our theory, $\kappa$ is as a property
of the filament connecting neighboring sites rather than as a
site-associated rigidity that connects next-nearest neighbor sites.
Following standard EM procedures, an arbitrary bond is either replaced
with probability $\P$ by a bond of stretching modulus $\mu$ and bending
rigidity $\kappa$ or removed with probability $1-\P$. The phonon Green's function after
this replacement is calculated as a perturbation with respect to the
uniform effective medium, treating the replaced bond as a scattering
potential $V$ on the EM Hamiltonian. The EMT self-consistency
condition requires that the disorder-averaged Green's function equals
that of the unperturbed EM, i.e., that the average $\TM$-matrix arising
from the perturbed bond vanishes, giving us equations determining $\mu_m$
and $\kappa_m$ for given $\P$.

In the EMT scattering potential $V$, the stretching term is simply
proportional to $\mu-\mu_m$ if the bond is occupied and $-\mu_m$ if it
is removed. The bending terms must, however, be treated differently
because replacing a bond generates two bending terms, both of which
involve second-neighbor interactions. This can be understood by
considering 4 sites $ijkl$ along a filament.  If one replaces bond
$jk$, two bending terms involving second-neighbors $ijk$ and $jkl$ are
generated in $V$.  The coefficients of these two bending terms can be
found by considering a composite filament connecting $ijkl$ that is
composed of rods with bending rigidity $\kappa_s$ between sites $jk$
and $\kappa_m$ between sites $ij$ and $jk$, respectively, where
$\kappa_s = \kappa$ if the bond is occupied and $\kappa_s=0$ if it is
removed.  A direct calculation of the minimum bending energy yields the
effective bending rigidity
\begin{eqnarray}\label{EQ:Vbeta}
    \kappa_c=2\Big(\frac{1}{\kappa_s}+\frac{1}{\kappa_m}\Big)^{-1},
\end{eqnarray}
and thus the coefficients of the two bending terms involving
$ijk$ and $jkl$ in $V$ is given by $\kappa_c-\kappa_m$.

To close the EMT self-consistency equation,
\begin{eqnarray}\label{EQ:SCET}
    \P \TM (\mu,\kappa) +(1-\P) \TM(0,0) =0 ,
\end{eqnarray}
where $\TM$ is the $\TM$-matrix constructed from the
perturbation of the scattering potential $V$, a third-neighbor
coupling $\lambda_m$
\begin{eqnarray}
	\frac{1}{2}\lambda_m \sum_{\langle ijkl \rangle} \left[\left( {\bf u}_{jk}-{\bf u}_{ij}\right)\times {\bf \hat{r}}_{ij}\right] \left[\left( {\bf u}_{kl}-{\bf u}_{jk}\right)\times {\bf \hat{r}}_{kl}\right]
\end{eqnarray}
must be introduced to the EM and to $V$ accordingly. Thus the EM is
characterized by 3 parameters $(\mu_m,\kappa_m,\lambda_m)$, determined
by the self-consistency equation~(\ref{EQ:SCET}).  We obtained
asymptotic solutions to the this equation for small $\kappa$ in the
vicinity of the CF isostatic point, in which $\mu_m$ can be written
into a scaling form same as that of Eq.~(\ref{eq:widomscale}) by identifying that the shear modulus $G=\sqrt{3}\mu_m/4$, and the scaling function is
\begin{eqnarray}
	\G_{\pm}(y) & \simeq & \frac{3}{2}\big(
		\pm 1+\sqrt{1+4\mathcal{A}y/9} \big) \nonumber\\
\end{eqnarray}
with $\mathcal{A}\simeq 2.413$.

For $\kappa/\mu\ll\vert\Delta\P\vert^{\phi_{\textrm{EMT}}}$, to leading
order, the value for $\mu_m$ reduces to $3\mu\vert\Delta\P\vert$ for
$\Delta\P>0$, and $\frac{\mathcal{A}}{3} \kappa\vert\Delta\P\vert^{-1}
$ for $\Delta\P<0$.  For
$\kappa/\mu\gg\vert\Delta\P\vert^{\phi_{\textrm{EMT}}}$ we get
$\mu_m\simeq\sqrt{\mathcal{A}}\,\mu^{1/2} \kappa^{1/2}$.  
These three scaling regimes correspond to three
different slopes $0,1,1/2$ in the $G\vert\Delta\P\vert^{-f}$ vs
$\kappa\vert \Delta \P\vert^{-\phi}$ plot, as shown in
Fig.~\ref{fig:widom}a.

Effective medium theories for bond-diluted lattices with central-force
springs are straightforward because the springs reside on an individual
bond. In contrast, EMTs for lattices with bending forces are less so
because bending forces reside on two bonds and dilution removes only a
single bond at a time. Our solution is to treat a given bond as a filament
segment with bending modulus $\kappa_s$. The effective lattice bending
modulus for neighboring bonds with respective bending moduli $\kappa_b$
and $\kappa_m$ is given by Eq.~(\ref{EQ:Vbeta}). This treatment allows
us to unambiguously remove one bond at a time. The resultant effective
theory necessarily includes bend-stretch coupling. A previous EMT
theory \cite{Das2007} treated the bending problem by removing two bonds
at a time. The result was a theory that lacks bend-stretch coupling
and predicted separate thresholds for the development of
non-vanishing $\mu_m$ and $\kappa_m$, which is inconsistent with both the 
numerical and analytical EMT results presented here.

\begin{figure} \begin{center}
%\vspace{-0.1in}
\includegraphics[width=\columnwidth]{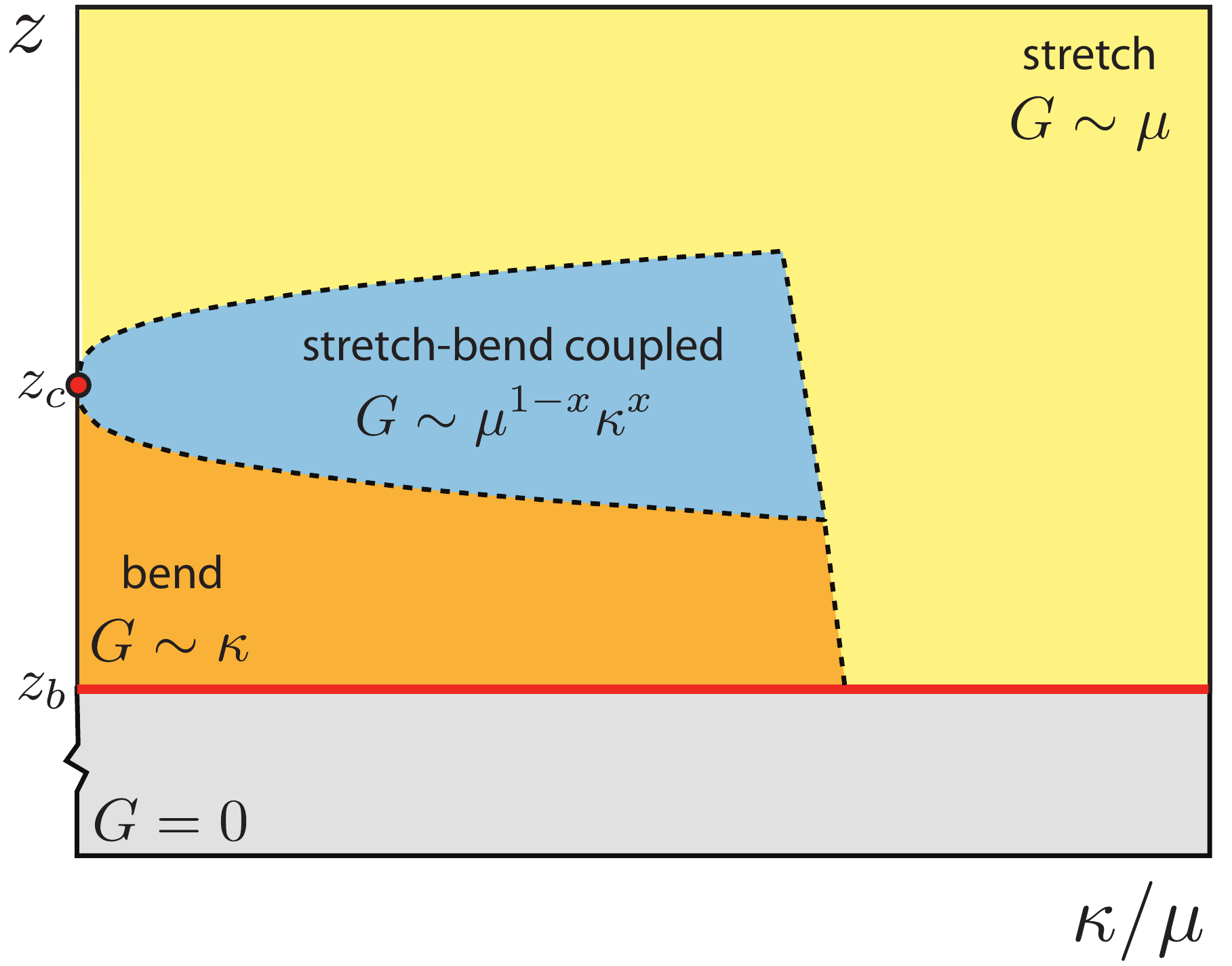}
\caption{
\label{fig:phasediagram} \textbf{Phase diagram} The phase diagram for diluted super-isostatic
networks. Above the rigidity percolation point $z_b$ there are three
distinct mechanical regimes: a stretching dominated regime with
$G\sim\mu$, a bending dominated regime with $G\sim\kappa$ and a regime
in which bend and stretch modes couple with $G\sim\mu^{1-x}\kappa^{x}$.
Here $x$ is related to the critical exponents $x=f/\phi$. We find here
that $x = 0.50\pm 0.01$ (2D triangular lattice) and $x=0.40\pm0.01$ (3D
FCC). The mechanical regimes are controlled by the isostatic point
$z_c$, which acts as a zero-temperature critical point.}
\end{center}\vspace{-0.2in} \end{figure}

\begin{acknowledgments}
This work was supported in part by NSF-DMR-0804900 (TCL and XM)
and in part by FOM/NWO (CPB and FCM). The authors thank M. Das and L. Jawerth 
for useful discussions.  

\emph{CPB and FCM designed the simulation model, which was
developed and executed by CPB. XM and TCL developed and
executed the EMT. All authors contributed to the writing of the
paper.}
\end{acknowledgments}

%\bibliographystyle{naturemag}
%\bibliography{isostaticity}

\end{document}